\documentclass[12pt,preprint]{aastex}

\shorttitle{GMOS-IFU observations of LV2}
\shortauthors{Vasconcelos et al.}

\newcommand{\mysp}{}\def\mysp/{}
\newcommand{\alloa}[3]{\ion{#1\mysp/}{#2}& #3}
\newcommand{\forba}[3]{[\ion{#1\mysp/}{#2}]& #3}

\begin{document}

\title{GMOS-IFU spectroscopy of 167-317 (LV2) proplyd in
Orion\altaffilmark{1}}

\author{M.J. Vasconcelos\altaffilmark{2}, A.H. Cerqueira\altaffilmark{2}
and H. Plana,}
\affil{LATO-DCET, Universidade Estadual de Santa Cruz, Rodovia
Ilh\'eus-Itabuna, km 16\\ Ilh\'eus, Bahia, Brazil - CEP 45662-000 \\
mjvasc@uesc.br, hoth@uesc.br, plana@uesc.br}

\author{A.C. Raga}
\affil{Instituto de Ciencias Nucleares, UNAM, Ap. Postal 70-543, CU,
D.F. 04510, M\'exico \\
raga@nucleares.unam.mx}

\and

\author{C. Morisset}
\affil{Instituto de Astronom{\'{\i}}a, UNAM, Ap. Postal 70-264, CU,
D.F. 04510, M\'exico \\
morisset@astroscu.unam.mx}

\altaffiltext{1}{Based on observations obtained at the Gemini Observatory,
which is operated by the Association of Universities for Research in
Astronomy, Inc., under a cooperative agreement with the NSF on behalf of
the Gemini partnership: the National Science Foundation (United States),
the Particle Physics and Astronomy Research Council (United Kingdom),
the National Research Council (Canada), CONICYT (Chile), the Australian
Research Council (Australia), CNPq (Brazil) and CONICET (Argentina).}

\altaffiltext{2}{Current Address: ICN-UNAM, Ap. Postal 70-543, CU,
D.F. 04510, M\'exico}

\begin{abstract}

We present high spatial resolution spectroscopic observations of the
proplyd 167-317 (LV2) near the Trapezium cluster in the Orion nebula,
obtained during the System Verification run of the Gemini Multi Object
Spectrograph (GMOS) Integral Field Unit (IFU) at the Gemini South
Observatory. We have detected 38 forbidden and permitted emission lines
associated with the proplyd and its redshifted jet. We have been able to
detect three velocity components in the profiles of some of these lines:
a peak with a 28-33 km s$^{-1}$ systemic velocity that is associated
with the photoevaporated proplyd flow, a highly redshifted component
associated with a previously reported jet (which has receding velocities
of about 80-120 km s$^{-1}$ with respect to the systemic velocity and
is spatially distributed to the southeast of the proplyd) and a less
obvious, approaching structure, which may possibly be associated with
a faint counter-jet with systemic velocity of $(-75 \pm 15)$ km s$^{-1}$. We
find evidences that the redshifted jet has a variable velocity, with slow
fluctuations as a function of the distance from the proplyd. We present
several background subtracted, spatially distributed emission line maps
and we use this information to obtain the dynamical characteristics
over the observed field. Using a simple model and with the extinction
corrected H$\alpha$ fluxes, we estimate the mass loss rate for both
the proplyd photoevaporated flow and the redshifted microjet, obtaining
$\dot{M}_{proplyd} = (6.2 \pm 0.6) \times 10^{-7}$ M$_{\sun}$ year$^{-1}$
and $\dot{M}_{jet} = (2.0 \pm 0.7) \times 10^{-8}$ M$_{\sun}$ year$^{-1}$,
respectively.

\end{abstract}

\keywords{ISM: clouds --- ISM: individual(\objectname[M 42]{Orion Nebula},
\object{LV2}, \object{167-317})}

\section{Introduction}

The Orion Nebula (M42) is the most active site of star formation of
the Orion Molecular Cloud and contains many young stars, among them the
Trapezium cluster. Its high mass young stellar members generate an intense
ultraviolet radiation field that photodissociates and photoionizes the
nearby material. These mechanisms promote the appearance of some complex
structures as the so-called proplyds \citep{odell94}. Proplyds
are low mass YSOs that are being exposed to an intense ultraviolet
radiation field which renders them visible. In Orion,
there are approximately 160 proplyds \citep{bally00} which are being
photoionized mainly by $\theta^1$ Ori C, an O6 spectral type star. There
are several objects that have been identified as being proplyds not only
near the Trapezium cluster \citep{bally00} but also in other star forming
regions \citep[like NGC 3372,][]{smith03}. Most of them share the same
features: a bow-shaped head that faces the ionization source, a tail, that
is primordially directed away from the source, a young star, that may (or
not) be visible, and a disk, sometimes seen in silhouette against the HII
region \citep[e.g.,][]{odell93,odell98,bally00,smith..05}. Early studies
of these objects, however, were only able to determine their apparent
ubiquity close to the $\theta^1$ Ori C star, their high ionization level
and, in later studies, the presence of high velocity structures (outflows)
associated with these condensations \citep[e.g.,][]{laques79,mea88,mea93}.

The proplyds are presently explained by a set of models which include
a photoevaporated wind, an ionization front and a photoionized wind
\citep{john98,storzer..98,henn98,bally98,odell98,rich..00}. \citet{john98}
proposed models in which far ultraviolet radiation (FUV) and extreme
ultraviolet radiation (EUV) are responsible for the mass loss rates of
the proplyds depending on the distance of the proplyd to the ionization
source. For the proplyds situated at intermediate distances from
the ionizing star, FUV photons penetrate the ionization front and
photodissociate and photoevaporate material from the accretion disk
surface, generating a supersonic neutral wind that passes through
a shock front before reaching the ionization front as a neutral,
subsonic wind. At the ionization front, EUV photons ionize the wind,
and the material is then reacelerated to supersonic velocities. The
interaction of this supersonic ionized wind with the star wind generates
bow shocks seen in H$\alpha$ and in [O III]$\lambda$5007 in some
proplyds \citep{bally98}. For proplyds closer to the ionizing star,
the ionization front reaches the disk surface, and the resulting wind
is initially subsonic, becoming supersonic further out.

In general, the models cited above are able to explain the main
observational features of the Orion proplyds. The region close to
the ionization front is responsible for most of the emission of these
objects. \citet{storzer..98} have shown that heating and dissociation of
H$_2$ can explain the observed [O I]$\lambda$6300 emission at the
disk surface \citep{bally98}. \citet{bally00} proposed that neutral gas
compressed between the shock front and the ionization front can explain
the [O III]$\lambda$5007 emission seen near the ionization front in
some objects. \citet{henn98} were able to reproduce the observed H$\alpha$
intensity profile of the proplyds in Orion in terms of accelerating
photoevaporated flows. Numerical 2D HD simulations with a treatment
of the radiative transfer \citep{rich..00} also reproduce the proplyd
morphology and emission, although in this work the treatment of the
diffuse radiation is somewhat simplified.

There are several open issues about these objects. One of them is related
with the calculation of the mass loss rate of the proplyds, that is
strongly model dependent and which may pose severe constraints on the
age of the Orion proplyds (as well as for $\theta^1$ Ori C). As the flux
of FUV photons dissociates the molecules in the disk of the proplyds,
the sub- or trans-sonic \emph{wind\/} is accelerated at the proplyd's
ionization front (IF) (see Henney et al. 2002 for a discussion), leading
to derived mass-loss rates of $\approx 8 \times 10^{-7} M_\odot$ yr$^{-1}$
which imply a short lifetime for these systems \citep{chur87}. It is hard
to reconcile such short lifetimes with the fact that the region of the
Trapezium cluster is populated by several proplyds. Improvements in both
models and observational techniques have slightly reduced the calculated
mass-loss rates for these systems, and \citet{henn02} find $\dot{M}
= 8.2 \times 10^{-7}M_\odot$ yr$^{-1}$ for the 167-317\footnote{The
proplyds mentioned in the present paper will be denoted following the
\citet{odell94} notation, that is based on the coordinate of the object.}
(LV2) proplyd, implying an age of less than $10^5$ yr for $\theta^1$
Ori C, and $\dot{M} \lesssim 1.5\times10^{-6}M_\odot$ yr$^{-1}$ for the
170-377, 177-341, 182-413, 244-440 proplyds \citep[e.g.,][]{henn99}.

It is presently known that many proplyds show, besides protostellar
features such as accretion disks and a young low mass protostar,
the presence of jets. \cite{bally00}, in a survey carried out using
the WFPC2 camera on the HST, find 23 objects which appear to have
collimated outflows seen as one-sided jets or bipolar chains of
bow shocks. \citet{mea02} also found kinematical traces of jets
associated with LV5 (158-323) and GMR 15 (161-307). The proplyd
167-317 shows evidence of the existence of a collimated outflow,
detected spectroscopically \citep{mea88,mea93,henn02}.  Until now, this
outflow was detected as a one-sided jet, with a P.A. of $120^{\circ}$
and a propagation velocity of about 100 km s$^{-1}$.  Recently,
\citet{smith..05} and \citet{bally05}, using the Wide Field Camera of
the Advanced Camera for Surveys (ACS/WFC) on HST, found jets associated
with silhouette disks in the outer regions of Orion.

In this work, we present the first Integral Field Unity (IFU) observation
of a proplyd. The observed object is 167-317, one of the brightest
proplyds of the Trapezium. We discuss one of the very first results
from the Gemini Multi-Object Spectrograph, in its IFU mode (hereafter,
GMOS-IFU), at the Gemini South Telescope.  In \S 2, we present the
observations and the steps to obtain a calibrated, clean datacube. In \S
3 we present the spectral analysis for the observed region of the 167-317
proplyd, as well as the observed lines and intensity maps.  We use the
observed line profiles to identify the outflows in the system. In \S 4,
we present the discussion and the conclusions.

\section{The observations and data reduction}

The data were taken during the System Verification run of the GMOS-IFU
at the Gemini South Telescope (GST), under the GS-2003B-SV-212 program,
on 2004 February 26$^{th}$ and 27$^{th}$. The science field of view
(FOV) is $3\farcs5 \times 5\arcsec$ with an array of 1000 lenslets
of $0\farcs2$. The sky is sampled with 500 lenses which are located
$1\arcmin$ from the science FOV.  We used the R831\_G5322 grating in
single slit mode, giving a sampling of 0.34\AA ~per pixel ($\approx 15$
km s$^{-1}$ per pixel at H$\alpha$) and a spectral coverage between
$\simeq 5515$ \AA~to $\simeq 7630$\AA. The spectral resolution (i.e.,
the instrumental profile) is 47 km s$^{-1}$ $\lesssim$ FWHM $\lesssim 63$
km s$^{-1}$ (the FWHM decreasing with increasing wavelength).  A field
centered on the 167-317 proplyd has been observed with an exposure time
of 300s.  A 60 s exposure of the standard star the Hiltner 600 has also been
taken in order to derive the sensitivity function.

The data were reduced using the standard Gemini IRAF v1.6 \footnote{IRAF
is distributed by the National Optical Astronomy Observatories,
which are operated by the Association of Universities for Research in
Astronomy, Inc., under cooperative agreement with the National Science
Foundation.} routines. An average bias image has been prepared using
the GBIAS task. The extraction has been worked out using the GCAL flat
and the response using the twilight flat. Spectra extraction has been
performed and wavelength calibration done using an arc taken during the
run. The 167-317 field and Hiltner 600 star have been processed in the
same way. The sensitivity function derived from the standard star has
been applied to the observed field.  Finally, data cubes have been built
using the GFCUBE routine. We maintain the IFU original resolution of $0
\farcs2$ px$^{-1}$ although an interpolation was performed in order to
turn the IFU hexagonal lens shape into squared pixels.

Because some of the emission lines are very intense, it was not possible
to eliminate cosmic rays using the standard GSCRREJ routine of the
IRAF Gemini v1.6 package. In order to remove cosmic rays we have then
programmed an IDL routine. This routine works directly on the data
cube by taking out impacts that are above a certain convenient level
calculated based on the mean intensity of nearby pixels. The cosmic
ray is then eliminated doing a linear interpolation in the wavelength
direction. With this procedure, it is possible to keep very intense lines
and eliminate lower level cosmic ray impacts that could be misinterpreted
as low emission lines.

Further manipulation of the cube, such as the production of channel
maps, emission line maps, dispersion and velocity maps, have been
done using Starlink \footnote{See documentation at this website:
http://star-www.rl.ac.uk/}, IDL routines and Fortran programs developed
specially for this purpose.

The subtraction of the background emission from the proplyd spectrum is
a very challenging task \citep{henn99}. This problem arises since the
background nebular emission is strongly inhomogeneous on an arc-second
scale. In this work, we have used standard $\chi^2$-method in order
to fit an intensity versus position plane in a semi-rectangular
field located $\approx 1\arcsec-1\farcs5 $ away from the peak proplyd
emission in each of the individual velocity channel maps.  We assure
that we only take samples of the background avoiding the region of the
proplyd. These planar fits to the nebular emission are then subtracted
from the corresponding velocity channel maps. In Figure \ref{f1}, we
show the observed field, the object orientation in the plane of the sky,
and also the region that defines our background (region 4, labeled
as R4 in Figure \ref{f1}). Region 4 (see Figure \ref{f1}) contains
76 spectra which are used to define the background. This Figure also
shows an internal box (labeled as R1, or region 1), for which we have
defined an arbitrary $xy$-coordinate system limited by $1 \le x \le 11$
(in the East direction) and $1 \le y \le 17$ (in the North direction; see
Figure \ref{f1}). In the following sections, we show maps of spatially
distributed physical variables that will be constrained to the limits
defined by this box. The spatial positions inside this domain will be
defined using the $xy$-coordinate system described above. Regions 2
and 3 (labelled as R2 and R3 in Figure \ref{f1}) define, respectively,
a region near the center of the 167-317 proplyd and a region which has
a high-velocity, redshifted feature (see below).

\section{Observational results: spectral line identification and high
velocity features} \label{sec:obs}

\subsection{The emission lines intensities and ratios}

The lines that we identify in the spectra have already been reported
in previous papers of the Orion Nebula \citep[e.g,.][]{baldwin00}.
We select 4 different regions to measure the line intensities, namely,
regions R1, R2, R3 and R4 (see Fig.~\ref{f1} and \S 2 for the coordinate
system definition).

Four spectra are then obtained by co-adding the spectra of the pixels
included in each region. The spectrum of region 1 is the result of the
sum of 187 pixels, for region 2 the sum of 6 pixels, region 3 represents
the spectrum of a single pixel and region 4 the sum of 76 pixels. A set
of 38 lines has been selected from a manual search for emission features
in the spectrum integrated over the selected fields. The intensities of the
emission lines are determined by fitting a flat continuum and integrating
over the whole emission feature \footnote{The radial velocity range that
defines the integration limits in each intensity line determination varies
from line to line, since the FWHM changes with the wavelength (see \S
2).}. Then, a mean flux was obtained dividing the integrated flux by the
total number of pixels of each region. Table~\ref{lines} gives the list
of the observed lines together with the mean flux of each region (after
subtraction of the emission from the background, region R4) and the ratio
to H$\alpha$ (normalized to H$\alpha=100$). The fourth and eigth columns
show the background emission and ratios to H$\alpha$, respectively, for
each line. The mean fluxes are given in units of $10^{-15}$ erg cm$^{-2}$
s$^{-1}$ px$^{-1}$, and are not corrected for reddening. The absolute
error of the mean flux is given in parentheses. This error was
calculated taking into account the variations of the local continuum.
The errors can be very large for the weaker lines.
Also, the detection of spectral lines in region R3 is affected by the low
signal to noise ratio. There are some lines which show stronger mean fluxes
in the background, namely, the lines of Si III $\lambda5740$,
Si II $\lambda5979$, [Ni II]$\lambda7378$ and [Fe II]$\lambda7155$
~(see Table \ref{lines}). These lines have low ionization potentials and
are expected to appear mainly in regions of lower ionization degree. The
lower line intensities within the proplyd indicate that these lines are
mainly emitted by the background nebula, and that they are absorbed at
least partially by the dust in the proplyd.

\begin{figure}
   \centering
   \includegraphics[width=15cm]{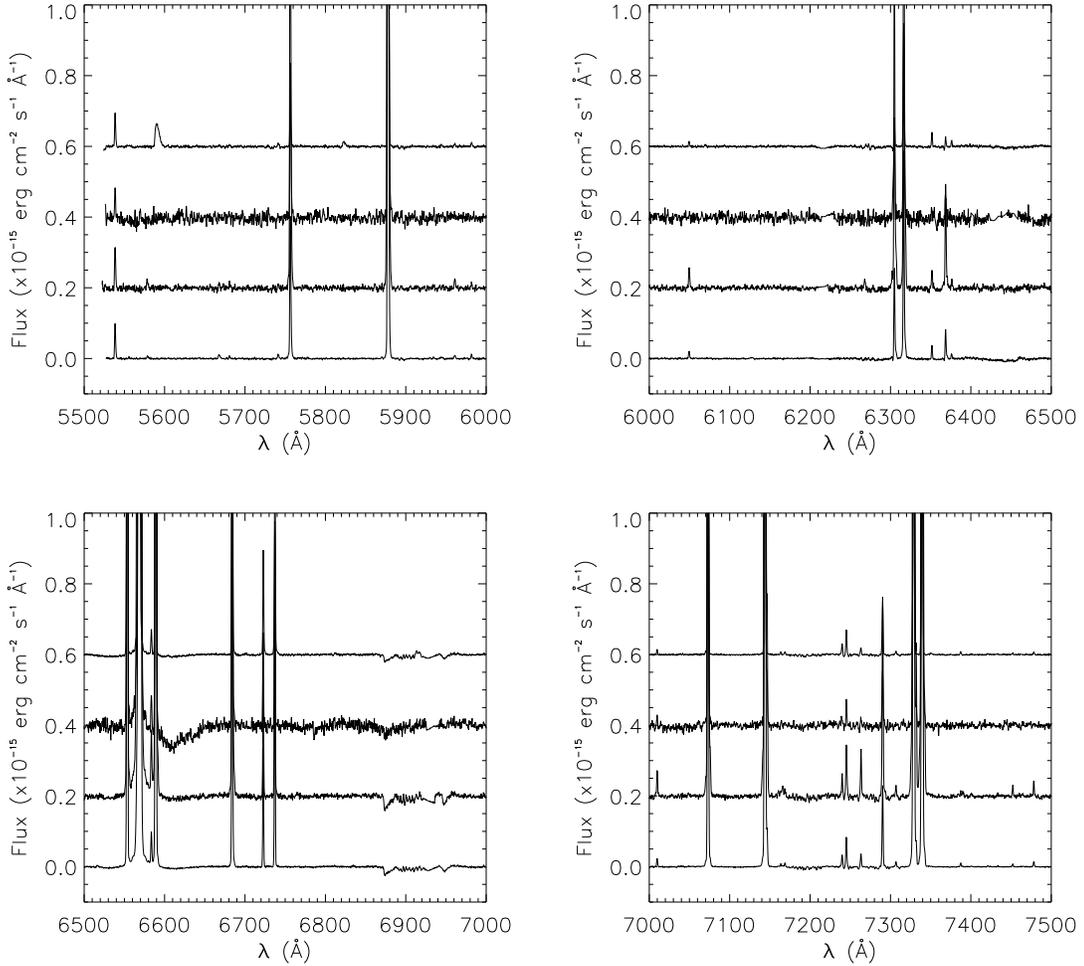}
   \caption{
   In each diagram, from bottom to top, are depicted, respectively, the 
   spectra for regions R1, R2, R3 and R4. A 3rd order cubic spline was
   fitted in order to subtract the continuum and a constant $n$ has been added 
   to the spectra of regions R2 ($n = 0.2$), R3 ($n = 0.3$) and R4 ($n = 
   0.4$) in order to produce an intensity offset between the successive
   spectra. Wavelengths ($\lambda$) are in \AA, and the fluxes are in 
   units of $10^{-15}$ erg cm$^{-2}$ s$^{-1}$ \AA$^{-1}$.
            }
              \label{f2}
   \end{figure}

In Figure \ref{f2}, we show the full extracted spectra from the data
cube, for regions R1 to R4 as defined in Figure \ref{f1}. In this figure,
the spectra for regions R1, R2, R3 and R4 are depicted from bottom
to top in each diagram. The data shown here are not background subtracted.
Most of the detected 38 lines can be clearly seen in the spectra. It is
also evident that the S/N is higher in regions R1 and R4
(which is due to the increase of the pixel number of these regions).  Here, a
third order cubic spline was fitted to the continuum and then subtracted
from the data. Because of the high order spline polynome subtraction, some minor
variations are still present near the H$\alpha$ line.

Figure \ref{f3} depicts (for region R1), 2D-intensity maps of the lines
[N II]$\lambda$5755, [N II]$\lambda$6548, H$\alpha$, [N II]$\lambda$6583,
[He I]$\lambda$7065 and [Ar III]$\lambda$7135, superimposed by line
profiles. The data shown here are background subtracted, as described
in \S 2.  The line profiles can be better seen in Figure \ref{f4}, in
which we show the mean line profile for each line and for each of the
regions R1, R2, R3 and R4, in the same diagram. As could be anticipated,
the flux is more intense in region R2 (dotted line), where the proplyd
is located, except for the [N II]$\lambda6548$ line, which shows a
stronger flux in region R3.  For the H$\alpha$ and [N II]$\lambda6583$
lines the emission of region R3 is more intense than the emission of
region R1. For lines with good S/N, the presence of a redshifted feature
starts to become clear. An analisys of the profile components will be
given in more detail below.

In order to see how the spectra of the different lines change spatially,
in Figure \ref{f5} we present an intensity {\it versus} position plot
for the same lines shown in Figures \ref{f3} and \ref{f4}. We show how
the intensity of these lines changes along a diagonal line, that crosses
region R1 passing through the pixel (3,4) (region R3) and through the
centre of region R2. We note that there are 2 intensity maxima
for the H$\alpha$, [N II]$\lambda 6583$ and [N II]$\lambda 6548$
lines, clearly showing in which lines region R3 is clearly visible
(also see Figure \ref{f3}). 

\subsection{High velocity features}

\begin{figure}
   \centering
   \vskip -2truecm
   \includegraphics[width=15cm]{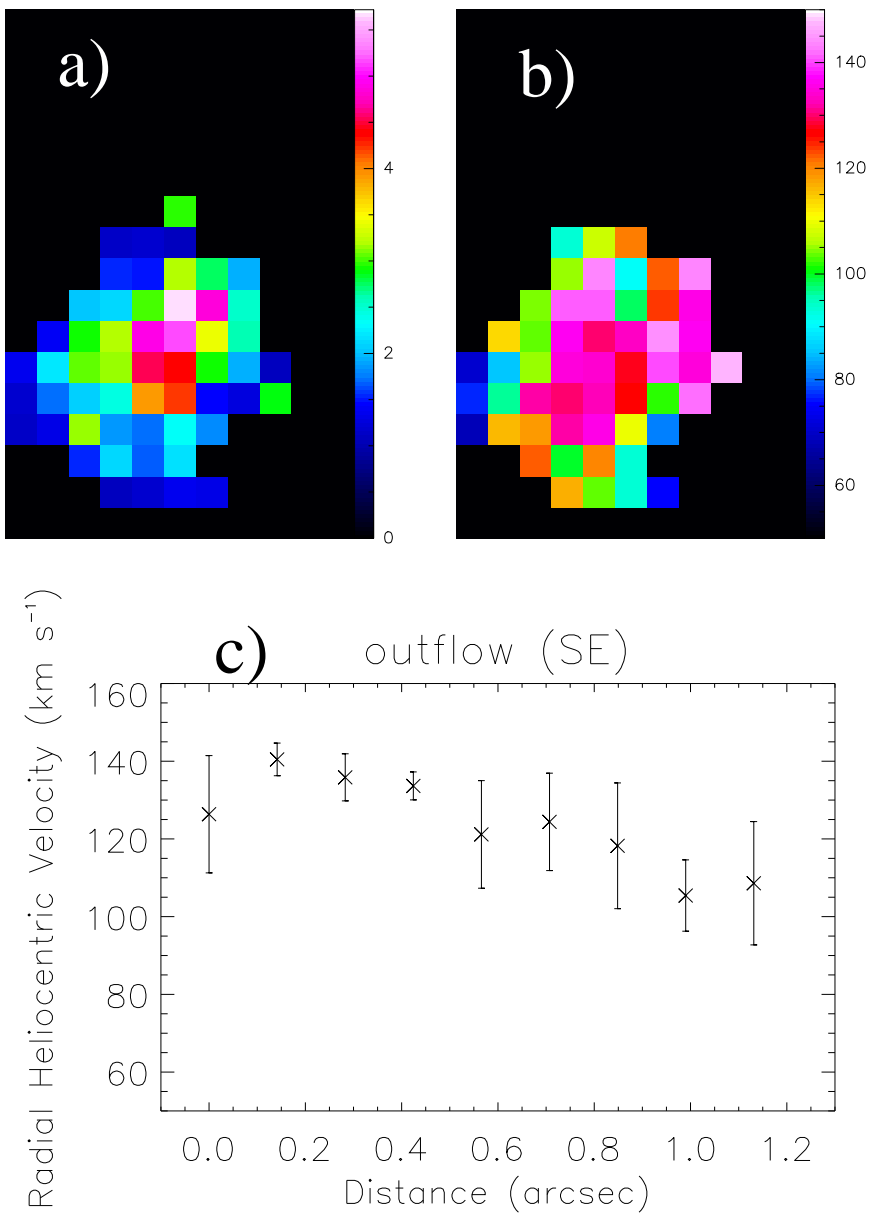}
   \vskip -6truecm
   \caption{
   a) The intensity, in units of $1\times10^{-15}$ erg cm$^{-2}$ s$^{-1}$
   px$^{-1}$, as a function of position in region R1. b) The velocity,
   in km s$^{-1}$ for the redshifted Gaussian fit to the H$\alpha$
   profiles. c) The jet velocity variation as a function of distance
   from the proplyd (see the text for a discussion). The velocities take
   into account the systemic radial velocity of $v_{\odot} \approx 26$
   km s$^{-1}$. The error associated with the radial velocity measurement
   is of the order of 2.5 km s$^{-1}$ (see the text for a discussion).
           }
              \label{f9}
   \end{figure}

We associate region R3 with the redshifted
jet of the proplyd, although in Figure \ref{f9} we will show that the
emission of the jet is not restricted only to this pixel.
A redshifted jet associated with the 167-317 proplyd was previously detected,
with a propagation velocity of $\approx$ 100 km s$^{-1}$
\citep[e.g.,][]{henn02}. We have also detected this high velocity feature
in our data. In order to see this, we examine the behaviour of the
moments of the radial velocity distribution. These are the flux-weighted
mean radial velocity $<v>$, the flux-weighted rms width of the line
$\Delta v^2$ and the skewness $\Delta v^3$, that are given,
respectively, by:

\begin{equation}\label{eq2}
 <v>=\frac{1}{I}\cdot \int I_vvdv
\end{equation}

\begin{equation}\label{eq3}
\Delta v^2=\frac{1}{I}\cdot\int I_v(v-<v>)^2dv 
\end{equation}

\begin{equation}\label{eq4}
\Delta v^3=\frac{1}{I}\cdot\int I_v(v-<v>)^3dv 
\end{equation}

\noindent where

\begin{equation}\label{eq1}
 I=\int I_vdv
\end{equation}

The integrated intensity (see eq. \ref{eq1} and Figure \ref{f3}) shows
that if we move away from the proplyd position (R2), the emission for
several lines (for example, H$\alpha$ and [Ar III]) rapidly drops to
low values (close to the background value).  We also note that the maps
for the flux-weighted mean radial velocity \footnote{Unless explicitly
mentioned, the radial velocities presented here are not corrected for the
systemic radial velocity, that is of the order of $v_{\odot} \approx 26$
km s$^{-1}$ \citep[e.g.,][]{mea93}. The error in the radial velocity
measurements is of the order of 2.5 km s$^{-1}$.} (not shown here;
see eq. \ref{eq2}) show that $<v> = 50 \pm 20$ km s$^{-1}$.

\begin{figure}
   \centering
   \vskip -2truecm
   \includegraphics[width=10cm]{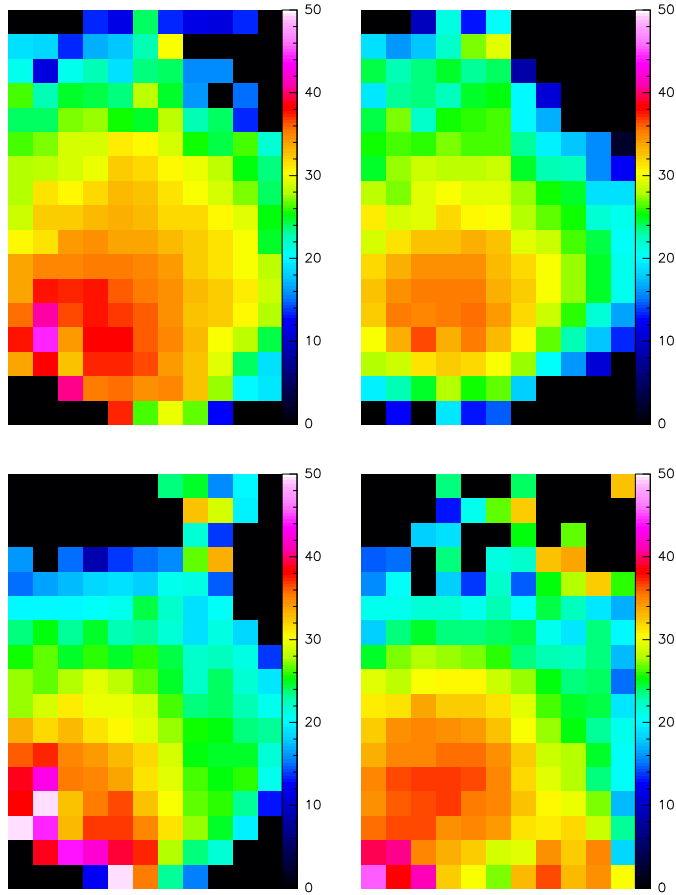}
   \vskip -4truecm
   \caption{
   The flux-weighted rms width (top) and skewnes (bottom) moments
   of the intensity vs. velocity distribution, as given by equations 
   (\ref{eq3}) and (\ref{eq4}), respectively, for the H$\alpha$ (left) 
   and [Ar III]$\lambda$7135.8 (right) lines. The grey scale on the
   right side of each map is in unit of km s$^{-1}$. We note that both
   integrals show positive values and peak in a SE region, indicating the 
   presence of a redshifted component towards the SE direction. The 
   velocities are not corrected by the systemic radial velocity, and the 
   integration limits in all maps are from $0 \le v_{rad} \le 200$ km 
   s$^{-1}$.  The white pixels have values close to zero.
             }
              \label{f6}
   \end{figure}

The flux-weighted rms width of the line $(\Delta v^2)^{1/2}$ and the
skewness $(\Delta v^3)^{1/3}$ moment of the radial velocity distribution
(see equations \ref{eq3} and \ref{eq4}, respectively) are depicted
in Figure \ref{f6}, for the H$\alpha$ (left) and [Ar III] (right)
lines. Both of them were computed after carrying out both background and
continuum subtractions. From Figure \ref{f6}, we see that there is a clear
enhancement of both of these moments from the proplyd position (R2 region)
towards the R3 region (i.e., towards the SE direction, at a PA $\approx
135^{\circ}$). This behaviour of the moments leads us to infer that: 1)
there is an increase in the line width as we go from the proplyd position
to the SE direction; and 2) that there are redshifted wings in the line
profiles. These results indicate the presence of a redshifted outflow
in this region. We note that the same behaviour is seen in the He I
$\lambda$6678.15, [N II]$\lambda$6583.46 lines (not shown here).

In order to confirm the presence of high velocity components in the
observed field, as well as to see the behaviour of the photoevaporated
proplyd flow and the background emission, we have computed three-component
Gaussian minimum $\chi^2$ fits for each position-dependent line profile. The
profiles of all emission lines have a major peak at $v_{rad} = 40-60$
km s$^{-1}$, the exact radial velocity of this peak changing with
spatial position.  In some spectral lines, there is an evident second
peak at redshifted velocities ranging from 100 to 150 km s$^{-1}$,
corresponding to the jet associated with the proplyd. Finally, we have
been able to detect a blueshifted component, which is fainter than the
redshifted components.  As an example of our three-component Gaussian
fit, in Figure \ref{f7} we show the data (full line) and the fit
(crosses), for the H$\alpha$ and [Ar III]$\lambda$7135 lines in the
($x$,$y$)=(6,7) position. In this figure, for each emission line,
the top-left panel represents the data and the three Gaussian fits,
the top-right panel shows the main, low velocity component (and the
Gaussian fit for this component), the bottom-left panel depicts the
data minus the main component together with the fits for the blue-
and red-shifted components and, finally, the bottom-right panel shows
the residual, obtained by subtracting the three-component fit from the
observed line profile. The fits depicted in this figure show the presence
of redshifted and blueshifted components. The redshifted component is
present in several lines in pixels around the SE direction, as already
mentioned before. On the other hand, blueshifted emission can be found
in several pixels to the NW of the proplyd peak emission as can be seen
in the left bottom panel of Figure \ref{f7} and this could represent
the first spectroscopic determination of the presence of a blueshifted
counter-jet, with systemic velocities of -60 km s$^{-1}$ $\lesssim v_{rad}
\lesssim -90$ km s$^{-1}$.  This blueshifted component is also suggested
by the fitted profiles of HeI $\lambda$7065 (not shown here).

Figure \ref{f8} depicts the spatially distributed intensity over
the R1 region (see Figure \ref{f1}), of the main peak for the [N
II]$\lambda$5754 (top-left), [N II]$\lambda$6548 (top-middle), H$\alpha$
(top-right), [N II]$\lambda$6583 (bottom-left),  HeI $\lambda$7035
(bottom-middle) and [Ar III]$\lambda$7135 (bottom-right) lines. The
values shown in this Figure were obtained from three-component Gaussian
fits to the observed line profile. All of the lines peak at the proplyd
position, at ($x$,$y$) = (6,7), inside the R2 region, and three of them
(namely [N II]$\lambda$6548, top-middle; H$\alpha$, top-right; and
[N II]$\lambda$6583, bottom-left), have a secondary, less intense peak
at ($x$,$y$)=(3,4) (i.e., in the R3 region; see Figure \ref{f1} for the
definition of the coordinate system). This secondary peak is probably related
to the jet, since, as we have seen before, the jet propagates from the R2
region towards the SE direction. It is also interesting to note that there
is a tail of faint emission (compared with the maximum in a given map)
that extends in the NE direction, and that can be seen in both of the
[N II] lines that bracket H$\alpha$ (top-middle and bottom-left maps in
Figure \ref{f8}).

In Figure \ref{f9}, we show the spatially dependent intensity
(Fig. \ref{f9}a) and the central velocity (Fig. \ref{f9}b) of the
redshifted component of the fitted H$\alpha$ profile.  We limit the maps
to the spatial pixels which have a skewness $\vert (\Delta v^3)^{1/3}
\vert > 30$ km s$^{-1}$. The pixels which satisfy this criterium have
a well defined high velocity, redshifted component. These figures show
that the high velocity component has an intensity maximum near the
center of the proplyd, with an extension towards the SE direction (along
the jet axis), surrounded by a region of decreasing
fluxes. The high intensity spike seen in H$\alpha$, [N II]$ \lambda6548$
and [N II]$ \lambda6583$ lines (Figure \ref{f3}) can be seen
here. The jet shows a spread in velocity values, although most of the
pixels present velocities around 120-140 km s$^{-1}$. This figure also
shows that the jet velocity decreases with increasing distance from the proplyd.
This trend can be seen in the Figure \ref{f9}c, where we
show the radial velocity of the redshifted component as a function of
distance from the proplyd position, ($x$,$y$) = (6,7) in region R1. To
build this figure, we assume that the jet is propagating in a PA $\approx
135^{\circ}$ position angle. We define a box with one axis aligned with
the jet axis, and the second, perpendicular dimension extending two pixels to
each side of the jet axis. We then take averages (of the mean velocity
of the redshifted component) perpendicular to the jet direction and plot
the resulting velocity as a function of distance from the proplyd
(see the bottom frame of Figure \ref{f8}).  There is an indication that
the jet velocity is slowly diminishing as a function of distance from the source.

\subsection{The spatial distribution of the nitrogen ratio}

In Figure \ref{f10}a, we show a map for the R1 region of the line ratio:

\begin{equation}
\frac{[{\rm N~II}]\lambda 6548 + [{\rm N~II}]\lambda 6583}{[{\rm
N~II}]\lambda 5754} = \frac{A_{6548} \sigma_{6548} + A_{6583}
\sigma_{6583}}{A_{5754} \sigma_{5754}},
\end{equation}

\noindent where $A$ and $\sigma$ are the height and the dispersion
of the fitted Gaussian profile (of the main, low velocity component).
The [NII](6548+6583)/5754  ratio is a classical electron temperature
diagnostic of low/medium density nebulae ($N_e < 10^{5}$ cm$^{-3}$). At
high densities, due to the collisional desexcitations, this ratio cannot be
used for the determination of the electron temperature, but it turns out to
be a quite good electron density diagnostic. This can be seen, for example,
from the \citet{oster} equation:

\begin{equation}
[NII](6548+6583)/5754 = \frac{6.91 \exp ( 2.5 \times 10^{4} / T_{e})}{1 +
2.5 \times 10^{-3} (N_{e} / T_{e}^{1/2})}
\end{equation}

Figure \ref{f10}b shows the lines corresponding to [NII](6548+6583)/5754
= 9, 10, 11, 80, 90, 100 in a $T_e - N_e$ plane. We can see that
for the values obtained in our case (line ratios close to 10), the
[NII](6548+6583)/5754 ratio is strongly dependent on the electron density,
and that the dependence on the electron temperature is small.

In Figures \ref{f10}c and \ref{f10}d, we show the densities computed
from the [NII](6548+6583)/5754 ratio for $T=10^4$ K and $T=1.5 \times
10^4$ K, respectively.  Although presenting a very complicated pattern,
these figures show that the region coinciding with the proplyd center,
and extending towards the SE direction (the jet propagation direction),
shows the lowest line ratio, indicating electron densities ($n_e$)
of the order of $2\times10^6$ cm$^{-3}$. This value is at least
in qualitatively agreement with the $n_0=(3.0 \pm 0.5) \times 10^6$
cm$^{-3}$ obtained by \cite{henn02} considering a $T=1.2\times 10^4$ K
temperature for the ionization front (that is inside the R2 region). We
can also note that both figures are similar, presenting few differences
in the density values.

The extended structure from the R2 region towards the SE direction is
surrounded by a region in which the nitrogen line ratio reaches values
of up to $\approx 30$. In this limit, $n_e < 10^5$ cm$^{-3}$. It is
interesting to note that in the clump associated with the redshifted jet,
the nitrogen line ratio increases substantially, indicating a decrease in
the electron density when compared with the same values near the center
of the proplyd. We have also obtained the [S II]$\lambda6716,\lambda6730$
emission maps. Unfortunately, we are not able to use these lines as a
diagnostic because neither the ratio of this doublet is a good electron
density indicator for this case (since it is constant for densities
higher than 10$^5$ cm$^{-3}$) nor do we have a good enough signal to
noise ratio for these lines.  Furthermore, the background subtracted
spectra for these two lines give us negative fluxes in the proplyd region
(see Table~\ref{lines}).  The presence of these negative features in
the line profiles is probably related with the presence of dust, as
discussed by \cite{henn99}.

\section{Discussion and conclusions}

\begin{figure}
   \centering
   \includegraphics[width=9cm]{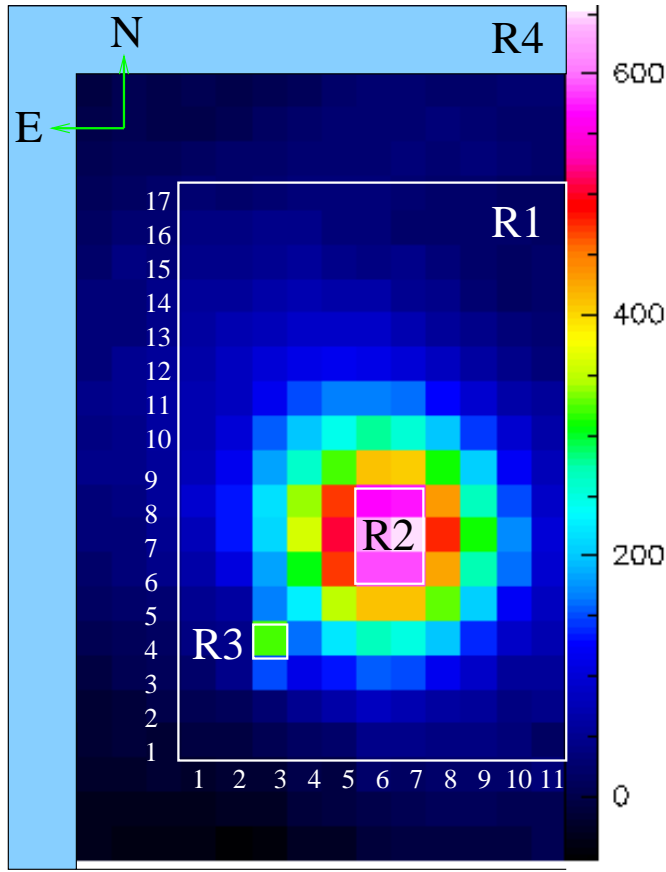}
   \caption{
    An H$\alpha$ integrated emission map (shown with a linear grey scale 
    in arbitrary units) showing the location of the 167-317 proplyd with 
    respect to the scientific field of view (SFOV), and its orientation: 
    North is up and East is left, as indicated by the arrows on the top-left
    corner of the figure. The field has 16 pixels in the East direction
    and 25 in the North direction. The R4 frame located at the left and 
    top parts of the figure represents the region choosen to define the
    background, which covers two columns of pixels on both the North and
    East directions. An internal box (R1), approximately centered on the 
    object emission defines a coordinate system, and the numbers on the
    bottom axis and on the left axis of this internal box are referred to
    in the text as the $x$- and $y$-coordinates, respectively. This
    spatial sub-sample of spectra will be used throughout the paper. The
    $\theta^1$ Ori C star is located at $7\farcs8$ from the proplyd,
    at a P.A. = 236$^{\circ}$. Four regions are defined in this figure,
    namely R1 (the internal box), R2 (the proplyd center), R3 (a clump
    associated with a redshifted jet) and R4 (the background). We will
    refer to these regions throughout the paper.
            }
              \label{f1}
   \end{figure}

\begin{figure}
   \centering
   \includegraphics[width=15cm]{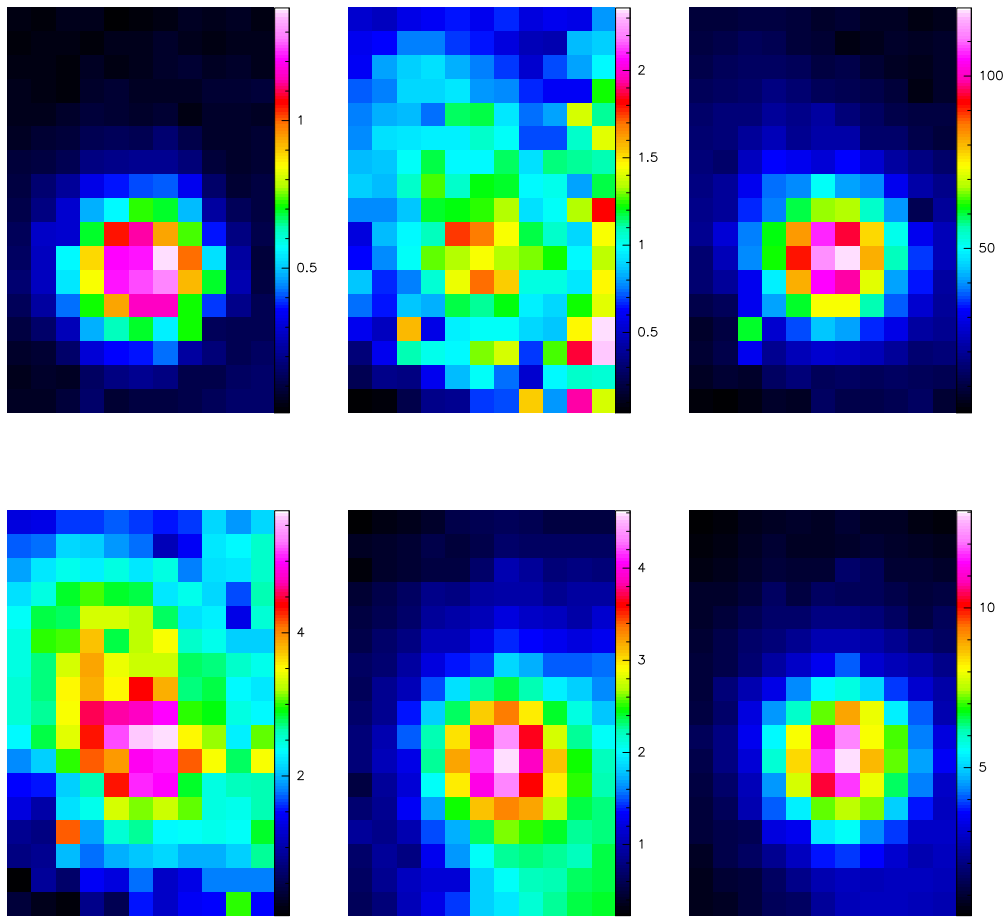}
   \vskip -9truecm
   \caption{
   The intensity of the main, low velocity peak (in units of $1 \times
   10^{-15}$ erg cm$^{-2}$ s$^{-1}$ px$^{-1}$) as a function of
   position for the [N II]$ \lambda$5754 (top-left), [N II] $\lambda$6548 
   (top-center), H$\alpha$ (top-right), [N II]$\lambda$6583
   (bottom-left), HeI $\lambda$6678 (bottom-center) and [Ar
   III]$\lambda$7135 (bottom-right) lines. Almost all of the lines
   peak at the proplyd position, and three of them, namely the [N
   II]$\lambda$6548 (top-centre), H$\alpha$ (top-right)  and [N
   II]$\lambda$6583 (bottom-left lines), have a secondary, less intense
   peak in region R3 (see Figure \ref{f1}).
           }
              \label{f8}
   \end{figure}

\begin{figure}
   \centering
   \includegraphics[width=16cm]{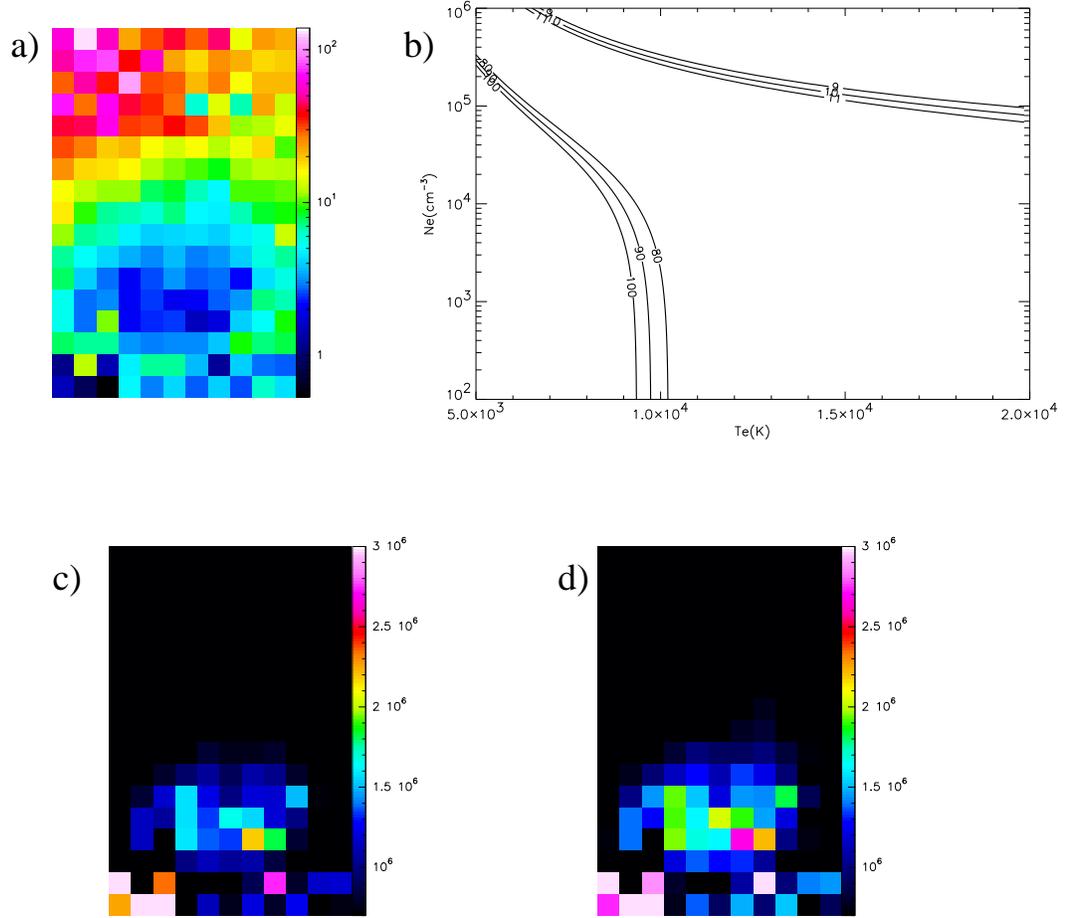}
   \vskip -8truecm
   \caption{$a)$ $N=$([NII]$\lambda6548$ + 
   [NII]$\lambda6583$)/[NII]$\lambda5754$ ratio map of region R1
   (see Figure \ref{f1}) of the observed field. $b)$  A [$T_e,n_e$]
   diagnostic diagram showing the locci corresponding to
   different values of the nitrogen line ratio. $c)$ The density 
   for an assumed temperature of $T=10^4$ K. $d)$ The density
   for an assumed temperature of $T=1.5\times10^4$ K.
            }
              \label{f10}%
   \end{figure}

We have presented in this work the first Integral Field Unit (IFU)
spectroscopic observations of a photoevaporating disk immersed in
a HII region. In particular, we have taken advantage of the System
Verification Run of the Gemini South Telescope Multi Object Spectrograph
(GMOS) to obtain spectra of the 167-317 proplyd in the Orion nebula. The
167-317 proplyd, also known as LV2 \citep[from the pioneering work from
][]{laques79}, is one of the brightest and best studied proplyds. In
a single exposure, we took 400 spectra with a spatial resolution of
$0\farcs2$. These spectra have been combined in order to optimize
this instrumental feature, and, before discussing of our results,
we want to make a few comments concerning the potential use of such
IFU data-cubes for these objects. As previously discussed in the
literature \citep{henn02}, the subtraction of the background emission
is a very challenging task. This problem arises since the background
near these proplyds (which are immersed in a highly non-homogeneous,
photoionized ambient medium) is very complex and changes in an arc-second
spatial scale. A careful computation of the background contribution
to the spectra of these objects is one of the most important tasks in
order to have reliable radial velocity measurements for both the
proplyd photoevaporated flow and the high velocity features (jets)
present in these systems.  Here, we have used the 76 spectra, defined
in Figure \ref{f1} as region R4, in order to obtain a planar
fit for the background emission. We think that such a definition of the
background emission is more precise, since we use several regions
to calculate a better function to describe the background. One more thing
to point out is that, with spatially distributed spectra, we are able
to confidently separate the contribution of the background from the
contribution of the object itself.

The data from the IFU observation of the 167-317 proplyd also allow us to
investigate the spatially dependent properties of the outflows associated
with this object. In particular, the 167-317 proplyd is known to have a
redshifted, collimated jet that propagates with heliocentric velocities
of $\approx 100$ km s$^{-1}$ towards the SE direction, and with a spatial
extension of $\sim 2\arcsec$ \citep[e.g, ][]{mea88, mea93, massey, henn00,
henn02}. We find that a prominent, high velocity redshifted component
can be detected in some emission lines, particularly in H$\alpha$, in
this SE direction. The redshifted jet has a trend in its radial velocity
(as a function of distance from the proplyd) with higher velocities
close to the proplyd and lower velocities at increasing distances.
We note that \cite{henn02} had previously suggested a variation in the
jet velocity as a function of distance from the source.

\begin{figure}
   \vskip -3truecm
   \centering
   \includegraphics[width=15cm]{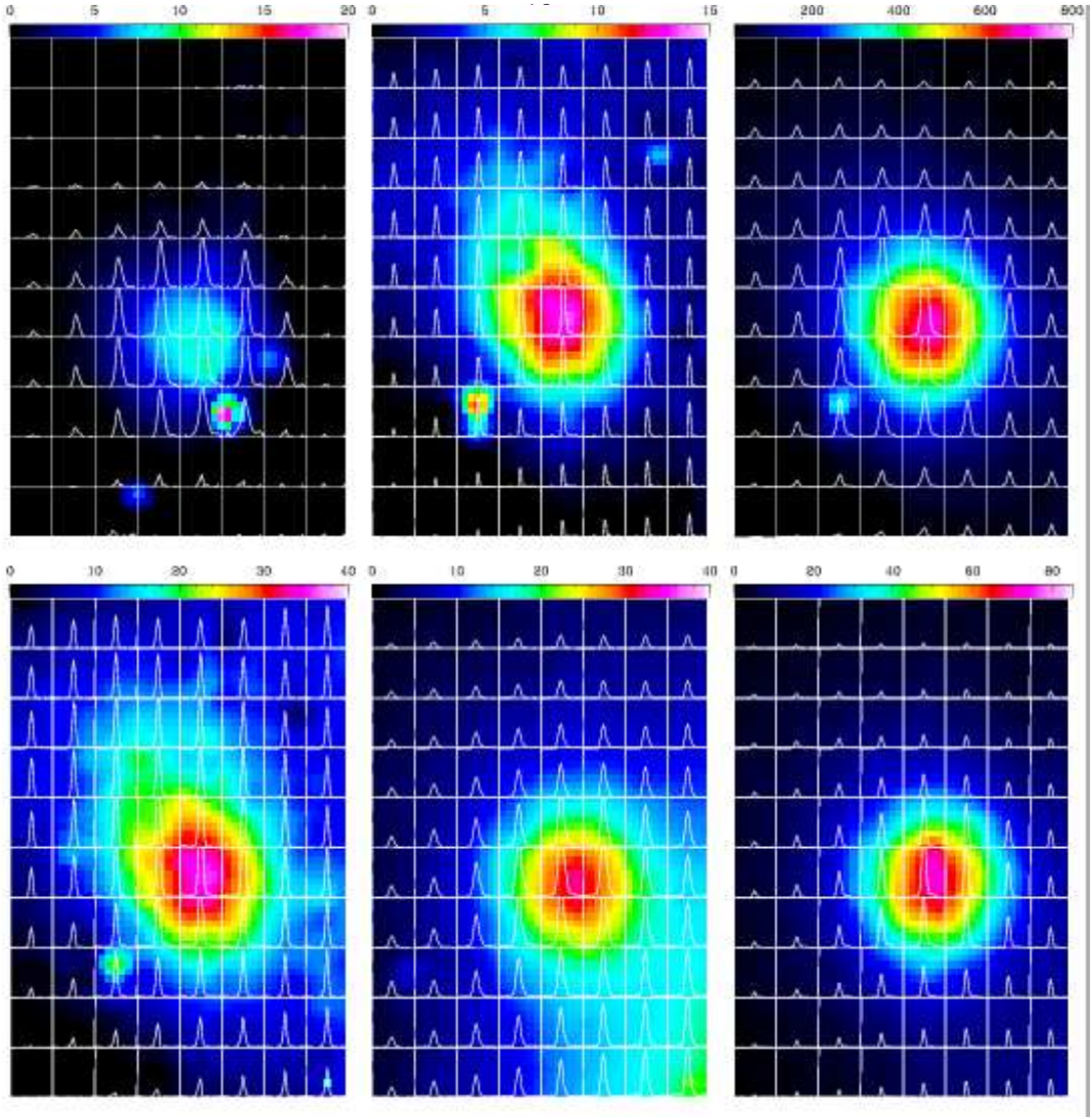}
   \caption{
   A sub-sample of the observed FOV, namely, the R1 region (see Figure
   \ref{f1}), showing the integrated flux (the grey-scale map) for the
   following lines: [N II]$\lambda$5755 (top-left), [N II]$\lambda$6548
   (top-center), H$\alpha$ (top-right), [N II]$\lambda$6583 (bottom-left),
   [He I]$\lambda$7065 (bottom-center) and [Ar III]$\lambda$7135
   (bottom-right). The linear bars give the grey-scale in unit of $1\times10^{-15}$ 
   erg cm$^{-2}$ s$^{-1}$ px$^{-1}$. The spectra in each position was 
   background subtracted. Superimposed on the intensity maps are the line 
   profiles for each box. See the text for a discussion. The orientation is 
   the same as in Figure \ref{f1} (i.e., North is up and East is left). The 
   original pixels (see Figure \ref{f1}) have been resampled in order to 
   smooth the emission features, and each box in these panels are slightly
   larger than the original pixels in Fig. \ref{f1}.
            }
              \label{f3}
   \end{figure}

\begin{figure}
   \vskip -3truecm
   \centering
   \includegraphics[width=15cm]{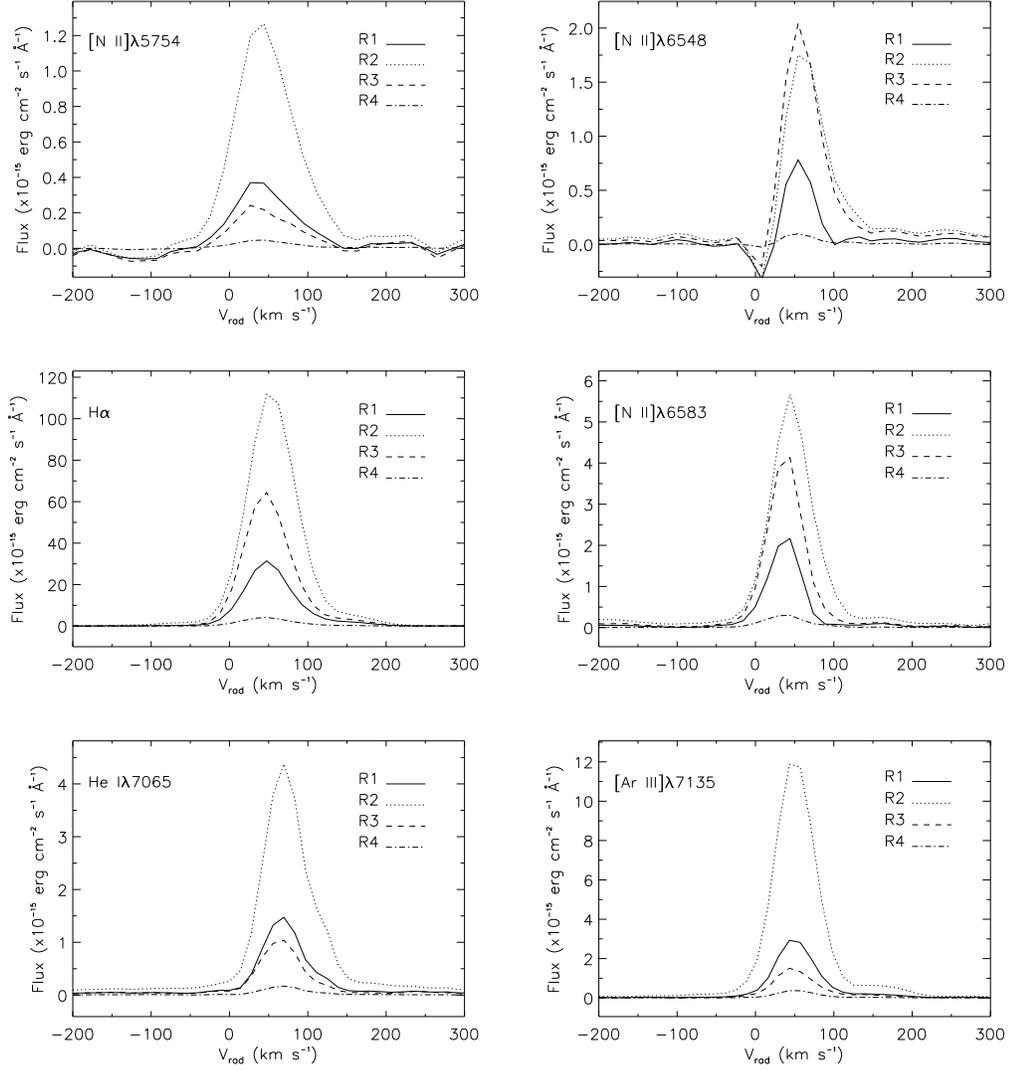}
   \vskip -5truecm
   \caption{
   An alternative view of the line profiles depicted in Figure \ref{f3}.
   Each diagram shows the line profile for regions R1 (continuous line),
   R2 (dotted line), R3 (dashed line) and R4 (dash-dotted line). The
   emission line is indicated on the top-left corner of each plot.
   The fluxes are in units of $1\times10^{-15}$ erg cm$^{-2}$ s$^{-1}$ per
   \AA ~ and tey are plotted against the radial velocity (in units of km 
   s$^{-1}$; not corrected for the heliocentric velocity).
            }
              \label{f4}
   \end{figure}

\begin{figure}
   \centering
   \includegraphics[width=15cm]{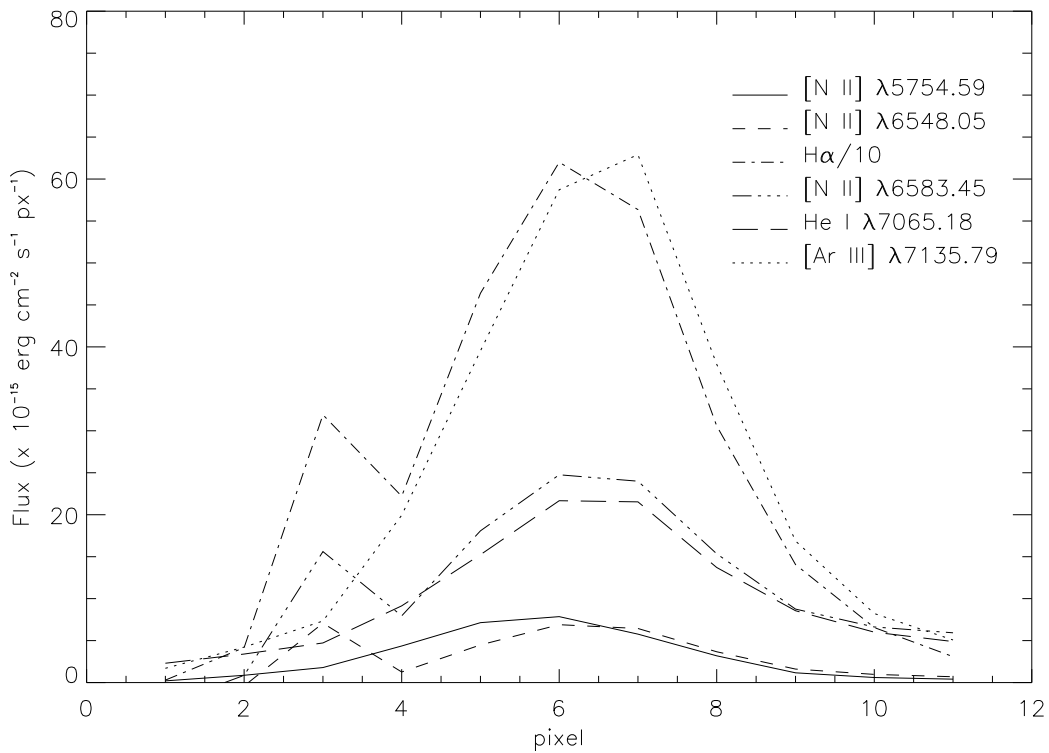}
   \caption{
   Intensity of the same lines depicted in Figure \ref{f3} as a function
   of the pixel position along a diagonal line that crosses region R1, 
   beginning at pixel (1,2), passing through region R3 (pixel 3,4) and 
   finishing at pixel (11,12). Each type of line shows the intensity of 
   a different spectral line: the solid line shows [N II] $\lambda 5754$; 
   the short dashed line, the intensity of [N II] $\lambda 6548$; the 
   dot-dashed line, H$\alpha$; the triple dotted-dashed line, [N II] 
   $\lambda 6583$; the long dashed line, He I $\lambda 7065$ and, finally, the
   dotted line shows [Ar III] $\lambda 7135$. In order to show all of the 
   intensities in the same plot, we have divided the H$\alpha$ intensity by a
   factor of 10.
           }
            \label{f5}
   \end{figure}

\begin{figure}
   \centering
   \includegraphics[width=12cm]{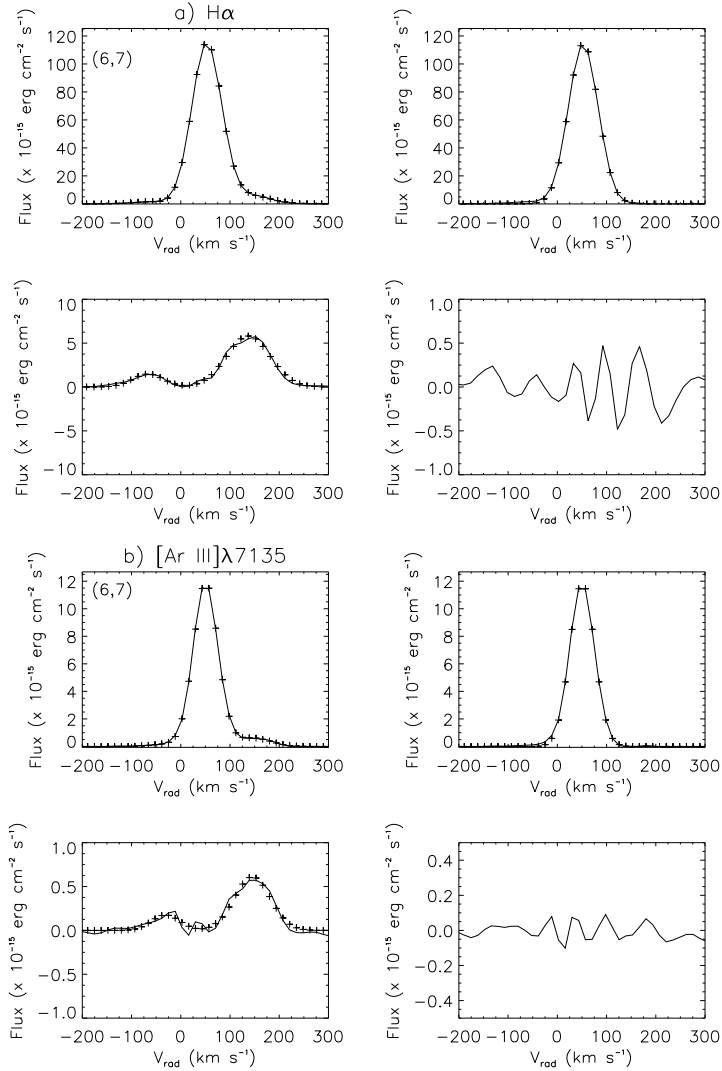}
   \vskip -5truecm
   \caption{
   a) Background and continuum subtracted data (full line) together
   with the Gaussian fit (crosses) for the H$\alpha$ line at position
   ($x$,$y$)=(6,7) (see Figure \ref{f1}). $Top-left:$ Line profile and
   the three Gaussian fit. $Top-right:$ Main, low velocity line profile
   and its Gaussian fit. $Bottom-left:$ The data profile subtracted
   from the main fitted profile, showing evidence for a blue- and a
   red-shifted component. $Bottom-right:$ The data profile subtracted
   from the three-component Gaussian fit. The residual profile is a
   random fluctuation around a flux of 0.5$\times 10^{-15}$ erg cm$^{-2}$
   s$^{-1}$. These fluctuations corresponds to 10\% of the flux for
   the blueshifted component, 0.5\% of the flux for the main
   component and 10\% of the flux for the redshifted component.
   b) The same but for the [Ar III]$\lambda$7135 line.
           }
              \label{f7}
   \end{figure}

There is a subtle peak (in several emission lines, particularly in
[N II]$\lambda$5754 and H$\alpha$) that is associated with the SE jet
emission, and located at $\approx 0\farcs67$ SE of the center of the
proplyd [at (x,y)=(3,4); the R3 region in Figure \ref{f1}].  For this
intensity peak, and also using the H$\alpha$ profiles of the neighbouring
pixels, we find a mean heliocentric velocity $v_{red} = 116 \pm 10$
km s$^{-1}$.  It is interesting to note that previous HST images and
spectroscopic analyses \citep{bally00, henn02} also reveal a spike in the
jet emission at $\approx 0\farcs4$ from the proplyd cusp. If we associate
these two emission regions as belonging to the same condensation, we can
estimate a proper motion of $\simeq 140$ km s$^{-1}$, which combined with
the $\simeq 116$ km s$^{-1}$ radial velocity give a full jet velocity of
$\approx 180 \pm 90$ km s$^{-1}$  (where the error was inferred from the
spatial resolution of our data). This velocity is of the same order as the
values inferred for Herbig-Haro jets associated with T Tauri stars. However,
the exact nature of this spike is unknown. We could relate this spike
with a bow shock of the jet, since the line profiles for both are similar
\citep{hartigan..87,beck..04}, showing a low velocity, more intense peak
together with a high velocity, less intense peak (similar to the line
profiles seen in Figures \ref{f3}, \ref{f4} and \ref{f7}).

We have found evidence for a blueshifted component in several emission
lines around the northwest tip of the proplyd position.  In particular,
the H$\alpha$ profiles reveal the presence of a blue-shifted component
with systemic velocity $v_{blue}= (-75 \pm 15)$ km s$^{-1}$, which
indicates that its velocity is similar though smaller than the one of
the redshifted jet. However, the intensities of the blue-shifted
features are much lower than the intensity of the redshifted components
associated with the SE jet: the counter-jet is at least four times
less intense than the redshifted jet (in H$\alpha$).  The presence of
such a blue-shifted component very close to the LV2 proplyd, together
with previously reported evidence that (at several arcseconds to the
NW of this proplyd; see Massey \& Meaburn 1995) there is a blueshifted
component (in the [O III]$\lambda$5007 line) strongly suggests the
existence of a faint counterjet.

It is interesting to estimate the mass loss rate for the proplyd
and the associated redshifted jet. In order to do this, we
consider a simple model that assumes that the proplyd flow arises from
a hemispherical, constant velocity wind that originates at the proplyd
ionization front, at a radius $r_0$.  Using the H$\alpha$ luminosity
$L_{H\alpha}$, we can then obtain the particle density $n_0$ at this
point and the mass loss rate, from:

\begin{equation}
n_0 = \left(\frac{2 L_{H\alpha}~{\mathrm e}^{\tau_{H\alpha}}} {4 \pi
\alpha_{H\alpha} h \nu_{H\alpha} r_0^3} \right)^{1/2},
\end{equation}

\begin{equation}
\dot{M} = 4 \pi r_0^2 n_0 \mu m_H c_{II}\nonumber,
\end{equation}
\noindent where ${\mathrm e}^{\tau_{H\alpha}}$ accounts for the extintion
correction, $\alpha_{H\alpha} = 5.83 \times 10^{-14}$ cm$^3$ is the
effective recombination coefficient for H$\alpha$, $h \nu_{H\alpha}$
is the energy of the H$\alpha$ transition, $\mu = 1.3$ is the mean
molecular weight and $c_{II} = 10$ km s$^{-1}$ is a typical value for the
sound velocity in an HII region. The extintion correction is obtained
from the base 10 logarithm of the extintion at $H\beta$ ($c_{H\beta}$)
which for the proplyd 167-317 is equal to 0.83 \citep{odell98} through
the relation, $c_{H\beta} = K \tau_{H\alpha}$ \citep{odell..92}, where
here we take $K$ = 0.56.  Assuming for the ionization front radius
a value $r_0=(7.9 \pm 0.2) \times 10^{14}$ cm \citep{henn02}
we obtain $n_0 = (2.3 \pm 0.6) \times 10^6$ cm$^{-3}$ and $\dot{M} =
(6.2 \pm 0.6) \times 10^{-7}$ M$_{\sun}$ year$^{-1}$. \cite{henn02}
have obtained a $\dot{M} = 8.2 \times 10^{-7}$ M$_{\sun}$ year$^{-1}$
$\pm 10\%$, and the small differences between both results may arise
from the different observational techniques employed in both cases.

For the redshifted jet, we assume that the outflow arises from the blob
located in region R3 (or pixel 3,4). In this case, the particle density,
mass and mass loss rate of the jet are given by,

\begin{equation}
n = \left(\frac{3 L_{H\alpha}}{4 \pi \alpha_{H\alpha} h\nu_{H\alpha}
r_{B}^3} \right)^{1/2},
\end{equation}

\begin{equation}
M = \frac{4 \pi}{3} r_B^3 n \mu m_H,
\end{equation}

\begin{equation}
\dot{M} \sim M \frac{v_j}{L_j},
\end{equation}
\noindent where $r_B = 6.7 \times 10^{14}$ cm is the radius of the
blob, $v_j = 180$ km s$^{-1}$ and $L_j = 4.5 \times 10^{15}$ cm are
the propagation velocity and length of the jet, respectively, derived
above. With these values, we obtain for the jet a mass loss rate equal to
$(2.0 \pm 0.7) \times 10^{-8}$ M$_{\sun}$ year$^{-1}$, which is similar
to the mass loss rate of typical HH jets from T Tauri stars.

We have been able to construct a ([NII]$\lambda6548$ +
[NII]$\lambda6583$)/[NII]$\lambda5754$ line ratio map. This ratio
indicates densities higher than $10^5$ cm$^{-3}$ for the proplyd emission
region, consistent with values previously reported in the literature.
We find a subtle enhancement of this ratio in the region of the redshifted
jet. Further observations of electron temperature and density diagnostic
lines would allow a more accurate determination of the spatial dependence
of these important parameters.

\acknowledgments

We thank the anonymous referee for his/her comments and suggestions,
which have improved significatively the presentation of the paper.
We are greatful to W. Henney for enlightning discussions.  M.J.V. and
A.H.C. would like to thank the staff of the Instituto de Ciencias
Nucleares, UNAM, M\'exico, for their kind hospitality, as well as for
partial financial support during our visit. We would like to thank Bryan
Miller, Tracy Beck and Rodrigo Carrasco from the Gemini Observatory,
and Bruno Castilho from LNA/CNPq, for help us in the early stages of
IFU-data processing. The work of M.J.V., A.H.C. and H.P. was partially
supported by the Bahia state funding agency (FAPESB), by the Universidade
Estadual de Santa Cruz (UESC) and by the Milenium Institute (CNPq
Project n. 62.0053/01-1-PADCT III/Mil\^enio). This work was partially
funded by the CONACYT grants 41320-F and 43103-F and the DGAPA (UNAM)
grants IN112602 and IN111803.

{}

\clearpage
\null
\vspace{3cm}
\begin{rotate}
\begin{table*}[ht]
\begin{scriptsize}
\begin{center}
\vspace{-3cm}
\caption{Detected emission lines for 167-317 \label{lines}}
\vspace{0.5cm}
\begin{tabular}{lr cccc  cccc}
\hline
\hline
\multicolumn{2}{c}{Line} & \multicolumn{4}{c}{Mean Flux ( $\times 10^{-15}
$ erg cm$^{-2}$ s$^{-1}$ px$^{-1}$)} & \multicolumn{4}{c}{Line ratio
(H$\alpha = 100$)}
\\
\hline
Ion & $\lambda$(\AA) & R1\tablenotemark{a} & R2 & R3 & R4 & R1 & R2 & R3 & R4 \\
\hline
\forba{Cl}{3}{5537.70}& -\tablenotemark{b} & 0.024 (0.007)\tablenotemark{c} & - & 0.155 (0.007) & - & 0.012 (0.003) & - & 0.164 (0.007) \\
\alloa{N}{2}{5666.63}& 0.024 (0.004) & 0.033 (0.003) & ND\tablenotemark{d} & 0.007 (0.003) & 0.058 (0.009) & 0.016 (0.002) &  ND & 0.007 (0.003) \\
\alloa{N}{2}{5679.56}& - & 0.002 (0.004) & ND & 0.012 (0.004) & - & 0.001 (0.002) & ND & 0.012 (0.004) \\
\alloa{Si}{3}{5739.73}& - & - & ND & 0.038 (0.002) & - & - & ND & 0.040 (0.002) \\
\forba{N}{2}{5754.59}& 0.535 (0.003) & 2.50 (0.01) & 0.638 (0.004) & 0.186 (0.003) & 1.283 (0.008) & 1.245 (0.005) & 0.598 (0.004) & 0.196 (0.003) \\
\alloa{He}{1}{5875.64}& 1.451 (0.003) & 7.80 (0.01) & 1.39 (0.03) & 4.327 (0.001) & 3.48 (0.01) & 3.879 (0.008) & 1.30 (0.03) & 4.568 (0.005) \\
\alloa{O}{1}{5958.60}& 0.007 (0.003) & 0.033 (0.007) & ND & 0.013 (0.003) & 0.016 (0.008) & 0.016 (0.004) & ND & 0.013 (0.003) \\
\alloa{Si}{2}{5978.93}& - & - & ND & 0.028 (0.002) & - & - & ND & 0.030 (0.002) \\
\alloa{O}{1}{6046.44}& 0.014 (0.002) & 0.09 (0.01) & ND & 0.020 (0.001) & 0.034 (0.004) & 0.045 (0.005) & ND & 0.021 (0.001) \\
\forba{O}{1}{6300.30}& 0.306 (0.004) & 1.58 (0.02) & 0.38 (0.02) & 0.086 (0.001) & 0.73 (0.01) & 0.78 (0.01) & 0.35 (0.02) & 0.091 (0.002) \\
\forba{S}{3}{6312.06}& 0.44 (0.02) & 2.11 (0.08) & 0.29 (0.04) & 0.516 (0.006) & 1.05 (0.04) & 1.05 (0.04) & 0.28 (0.04) & 0.545 (0.006) \\
\alloa{Si}{2}{6347.11}& - & 0.03 (0.01) & ND & 0.063 (0.004) & - & 0.015 (0.005) & ND & 0.066 (0.004) \\
\forba{O}{1}{6363.78}& 0.09 (0.01) & 0.51 (0.02) & 0.03 (0.03) & 0.03 (0.01) & 0.21 (0.03) & 0.25 (0.01) & 0.03 (0.02) & 0.03 (0.01) \\
\alloa{Si}{2}{6371.37}& - & 0.03 (0.01) & ND & 0.026 (0.003) & - & 0.018 (0.007) & ND & 0.027 (0.003) \\
\forba{N}{2}{6548.05}& 1.00 (0.01) & 2.55 (0.02) & 2.53 (0.04) & 3.475 (0.009) &2.39 (0.03) & 1.27 (0.01) & 2.34 (0.04) & 3.67 (0.01) \\
\alloa{H}{1}{6562.82}& 41.73 (0.02) & 201.20 (0.05) & 106.66 (0.02) & 94.73 (0.01) & 100.0 (0.4) & 100.0 (0.1) & 100.0 (0.1) & 100.0 (0.1) \\
\alloa{C}{2}{6578.05}& 0.02 (0.01) & 0.10 (0.02) & 0.05 (0.01) & 0.09 (0.01) & 0.04 (0.03) & 0.05 (0.01) & 0.05 (0.01) & 0.09 (0.01) \\
\forba{N}{2}{6583.45}& 3.134 (0.005) & 8.48 (0.02) & 5.304 (0.007) & 10.710 (0.001) & 7.51 (0.03) & 4.21 (0.01) & 4.972 (0.009) & 11.30 (0.01) \\
\alloa{He}{1}{6678.15}& 0.530 (0.007) & 2.621 (0.009) & 0.42 (0.02) & 1.241 (0.007) & 1.27 (0.02) & 1.303 (0.005) & 0.40 (0.02) & 1.310 (0.008) \\
\forba{S}{2}{6716.44}& 0.016 (0.005) & 0.029 (0.006) & - & 0.425 (0.003) & 0.04 (0.01) & 0.015 (0.003) & - & 0.448 (0.004) \\
\forba{S}{2}{6730.82}& 0.044 (0.002) & 0.06 (0.02) & - & 0.945 (0.002) & 0.107 (0.006) & 0.032 (0.008) & - & 0.998 (0.002) \\
\alloa{He}{1}{6989.48}& - & 0.016 (0.002) & ND & 0.005 (0.000)\tablenotemark{e} & - & 0.008 (0.001) & ND & 0.005 (0.000)\tablenotemark{e} \\
\alloa{O}{1}{7002.12}& 0.014 (0.004) & 0.088 (0.005) & 0.01 (0.01) & 0.022 (0.001) & 0.03 (0.01) & 0.044 (0.003) & 0.01 (0.01) & 0.023 (0.001) \\
\alloa{He}{1}{7065.18}& 0.995 (0.004) & 5.393 (0.009) & 0.39 (0.03) & 2.439 (0.001) & 2.38 (0.01) & 2.680 (0.005) & 0.37 (0.03) & 2.574 (0.003) \\
\forba{Ar}{3}{7135.79}& 3.713 (0.004) & 19.39 (0.02) & 1.88 (0.01) & 5.277 (0.003) & 8.90 (0.04) & 9.64 (0.01) & 1.76 (0.01) & 5.570 (0.007) \\
\forba{Fe}{2}{7155.16}& - & 0.06 (0.03) & ND & 0.005 (0.005) & - & 0.03 (0.01) & ND & 0.005 (0.006) \\
\alloa{He}{1}{7160.60}& 0.003 (0.004) & 0.007 (0.008) & ND & 0.017 (0.004) & 0.006 (0.009) & 0.004 (0.004) & ND & 0.018 (0.004) \\
\alloa{C}{2}{7231.33}& 0.004 (0.005) & 0.04 (0.01) & ND & 0.046 (0.004) & 0.01 (0.01) & 0.019 (0.005) & ND & 0.049 (0.004) \\
\forba{C}{2}{7236.42}& 0.024 (0.004) & 0.140 (0.005) & - & 0.108 (0.004) & 0.06 (0.01) & 0.070 (0.002) & - & 0.115 (0.004) \\
\alloa{O}{1}{7254.45}& 0.028 (0.001) & 0.17 (0.01) & ND & 0.027 (0.001) & 0.066 (0.003) & 0.086 (0.007) & ND & 0.028 (0.001) \\
\alloa{He}{1}{7281.35}& 0.13 (0.01) & 0.73 (0.02) & 0.04 (0.04) & 0.235 (0.007) & 0.32 (0.02) & 0.365 (0.008) & 0.03 (0.03) & 0.248 (0.008) \\
\alloa{He}{1}{7298.05}& 0.006 (0.004) & 0.040 (0.007) & ND & 0.020 (0.001) & 0.01 (0.01) & 0.020 (0.003) & ND & 0.021 (0.001) \\
\forba{O}{2}{7319.99}& 3.94 (0.04) & 21.9 (0.2) & 2.032 (0.008) & 1.488 (0.004) & 9.4 (0.1) & 10.89 (0.09) & 1.905 (0.008) & 1.571 (0.005) \\
\forba{O}{2}{7330.30}& 2.99 (0.08) & 16.38 (0.08) & 1.45 (0.08) & 1.12 (0.08) & 7.2 (0.2) & 8.14 (0.04) & 1.36 (0.07) & 1.18 (0.08) \\
\forba{Ni}{2}{7377.83}& 0.003 (0.001) & ND & ND & 0.013 (0.000)\tablenotemark{e} & 0.007 (0.002) & ND & ND & 0.014 (0.000)\tablenotemark{e} \\
\forba{N}{1}{7442.00}& 0.005 (0.003) & 0.024 (0.009) & ND & 0.008 (0.003) & 0.012 (0.008) & 0.012 (0.004) & ND & 0.008 (0.004) \\
\forba{N}{1}{7468.31}& 0.009 (0.002) & 0.049 (0.007) & ND & 0.011 (0.001) & 0.022 (0.004) & 0.024 (0.003) & ND & 0.012 (0.001) \\
\forba{C}{2}{7530.00}& 0.012 (0.000)\tablenotemark{e} & 0.014 (0.001) & ND & 0.010 (0.000)\tablenotemark{e} & 0.030 (0.001) & 0.007 (0.000)\tablenotemark{e} & ND & 0.011 (0.000)\tablenotemark{e} \\
\hline
\end{tabular}
\tablenotetext{a} {\ The regions R1, R2, R3 and R4 are defined in Figure 
\ref{f1}.}
\tablenotetext{b} {\ The symbol - implies that $I_{\rm region}-I_{\rm 
background} < 0$.}
\tablenotetext{c} {\ The terms inside the parentheses are the absolute errors
of the intensity of each line, for each one of the regions.}
\tablenotetext{d} {\ The symbol ND implies that the line were not detected in
that region.}
\tablenotetext{e} {\ Absolute error smaller than $6 \times 10^{-4}$.}
\end{center}
\end{scriptsize}
\end{table*}
\end{rotate}
\clearpage

%

%

%

\clearpage
\clearpage

\clearpage

\clearpage

\clearpage

%
\end{document}